\documentclass[conference]{IEEEtran}
\IEEEoverridecommandlockouts
\usepackage{cite}
\usepackage{float}
\usepackage[colorlinks,bookmarksopen,bookmarksnumbered,citecolor=blue, linkcolor=red, urlcolor=blue]{hyperref}
\usepackage{amsmath,amssymb,amsfonts}
\usepackage{algorithmic}
\usepackage{graphicx}
\usepackage{array}
\usepackage{amssymb}
\usepackage{adjustbox}
\usepackage{textcomp}
\usepackage{xcolor}
\usepackage{multirow} 
\usepackage{orcidlink}
\def\BibTeX{{\rm B\kern-.05em{\sc i\kern-.025em b}\kern-.08em
    T\kern-.1667em\lower.7ex\hbox{E}\kern-.125emX}}
\begin{document}

\title{Spatial-Frequency Dual Domain Attention Network For Medical Image Segmentation\\
\thanks{The corresponding authors: Xueshuo Xie* and Tao Li*}
}

\author{
\IEEEauthorblockN{1\textsuperscript{st} Zhenhuan Zhou\orcidlink{0009-0000-2187-7184}}
\IEEEauthorblockA{\textit{College of Computer Science} \\
\textit{Nankai University}\\
Tianjin, China \\
zhouzhenhuan@mail.nankai.edu.cn}
\\
\IEEEauthorblockN{4\textsuperscript{th} Rui Yao\orcidlink{0009-0008-0949-2247}}
\IEEEauthorblockA{\textit{The Department of Pediatric Dentistry} \\
\textit{Tianjin stomatological hospital}\\
Tianjin, China \\
yaorui73@163.com}
\and

\IEEEauthorblockN{2\textsuperscript{nd} Along He\orcidlink{0000-0003-1356-8757}}
\IEEEauthorblockA{\textit{College of Computer Science} \\
\textit{Nankai University}\\
Tianjin, China \\
healong2020@163.com}
\\
\IEEEauthorblockN{5\textsuperscript{th} Xueshuo Xie*\orcidlink{0000-0002-8245-8415}}
\IEEEauthorblockA{\textit{Haihe Lab of ITAI} \\
Tianjin, China \\
xueshuoxie@nankai.edu.cn}
\and

\IEEEauthorblockN{3\textsuperscript{rd} Yanlin Wu\orcidlink{0000-0002-2087-275X}}
\IEEEauthorblockA{\textit{College of Computer Science} \\
\textit{Nankai University}\\
Tianjin, China \\
1229145158@qq.com}
\\
\IEEEauthorblockN{6\textsuperscript{th} Tao Li*\orcidlink{0000-0002-1273-0487}}
\IEEEauthorblockA{\textit{College of Computer Science} \\
\textit{Nankai University and}\\
\textit{Haihe Lab of ITAI}\\
Tianjin, China \\
litao@nankai.edu.cn}

}

\maketitle

\begin{abstract}
In medical images, various types of lesions often manifest significant differences in their shape and texture. The majority of medical image segmentation networks exclusively learn features in the spatial domain, disregarding the abundant global information in the frequency domain. This results in a bias towards low-frequency components, neglecting crucial high-frequency information. To address these problems, we introduce SF-UNet, a spatial-frequency dual-domain attention network. It comprises two main components: the Multi-scale Progressive Channel Attention (MPCA) block, which progressively extract multi-scale features across adjacent encoder layers, and the lightweight Frequency-Spatial Attention (FSA) block, with only 0.05M parameters, enabling concurrent learning of texture and boundary features from both spatial and frequency domains. We validate the effectiveness of the proposed SF-UNet on three public datasets. Experimental results show that compared to previous state-of-the-art (SOTA) medical image segmentation networks. Codes will be released at \href{https://github.com/nkicsl/SF-UNet}{https://github.com/nkicsl/SF-UNet}.
\end{abstract}

\begin{IEEEkeywords}
Medical Image Segmentation, Multi-scale Feature, Frequency Domain Attention, CNN, Deep Learning
\end{IEEEkeywords}

\section{Introduction}

Convolutional Neural Networks (CNN) have been widely used in computer vision tasks due to their powerful feature extraction capabilities. In medical image segmentation, UNet \cite{ronneberger2015u} and its variants \cite{qin2023aia} \cite{wang2023lde} are representative deep learning networks and have achieved remarkable performances. However, such CNN segmentation models still have some limitations: Firstly, due to the limitation of kernel size, CNN often have a limited receptive field, lacking the ability to learn global information and long-range relationships. Secondly, in medical image segmentation, the multi-scale problem is a significant challenge because different types of lesions often exhibit large differences in shape and volume \cite{he2022progressive}. CNN-based encoders often gradually downsample the resolution of feature maps, which may lead to the neglect of some small lesions, causing semantic loss and ultimately affecting the reliability of segmentation results.


Transformer was originally proposed in natural language processing\cite{vaswani2017attention} and quickly applied to computer vision\cite{dosovitskiy2020image}. For medical images, global information and long-range relationship are crucial \cite{shamshad2023transformers}. However, transformers are not without their flaws. On one hand, global self-attention has quadratic computational complexity, leading to larger computational overhead and data requirements compared to CNN \cite{liu2021swin}. On the other hand, some researches have shown that global self-attention primarily focuses on the low-frequency components of the image, potentially neglecting high-frequency information to some extent \cite{zhang2023fsanet}. 

To solve the problems mentioned above, we proposed a U-shaped Spatial-Frequency Dual Domain Attention Network, named SF-UNet. It consists of two main blocks: the Multi-scale Progressive Channel Attention (MPCA) block progressively fuses features from adjacent levels of the encoder to generate cross-scale channel attention maps, enhancing the network's ability to learn multi-scale features and avoiding feature redundancy. The Frequency-Spatial Attention (FSA) block consists of two branches. One of them learns attention maps in the spatial domain, and the other branch applies 2D Discrete Fourier Transformation to the spatial feature maps, and then separating the frequency feature maps into high-frequency and low-frequency components. In the low-frequency component, we use a learnable filter to adaptively adjust the weights of each frequency component, enabling global feature learning from a holistic perspective. Finally, we restore the original high-frequency components of the feature maps to the filtered low-frequency components, aiming to preserve the integrity of high-frequency details in the decoders. We validated the proposed method on three public datasets, and the results demonstrate that SF-UNet can outperform previous SOTA medical image segmentation models.

	
	

\section{Method}
\subsection{Overall Structure}

\begin{figure*}[t]
\centerline{\includegraphics[width=\linewidth]{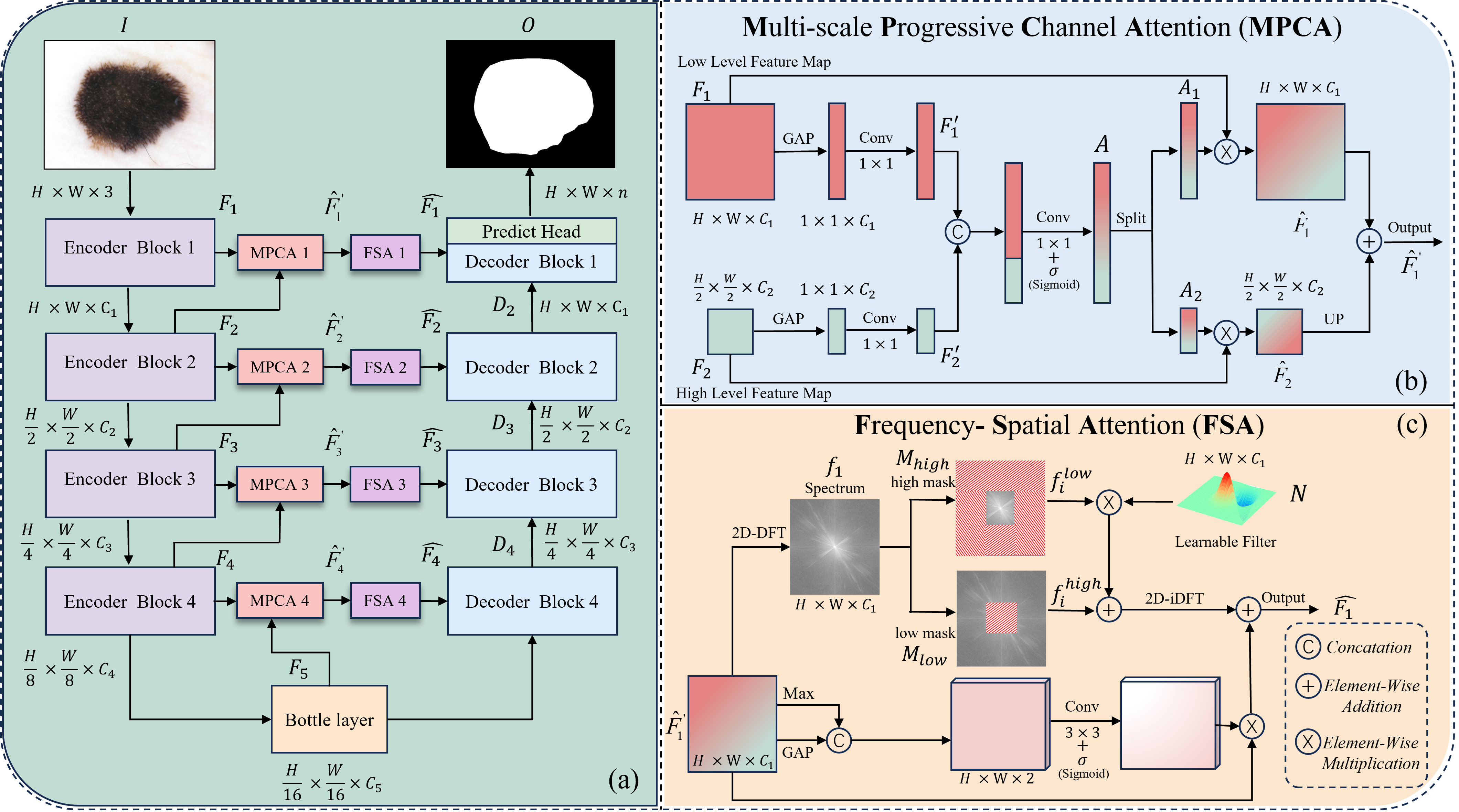}}
\vspace{-0.2cm}\caption{The overall architecture of the proposed network and detailed structures of each block. (a) the overall structure of SF-UNet, (b) the structure of MPCA. (c) the structure of FSA. 2D-DFT stands for 2D Discrete Fourier Transform, and 2D-iDFT stands for 2D Inverse Discrete Fourier Transform.}
\label{fig2}
\end{figure*}

The overall pipeline of SF-UNet is illustrated in Fig \ref{fig2}. The encoder of SF-UNet consists of four encoder blocks followed by a bottleneck layer, following the implementation of VGG16 \cite{simonyan2014very}. The encoder and decoder in SF-UNet are accompanied by MPCA and FSA blocks. Specifically, given an image $I \in \mathbb{R}^{H\times W \times C}$, H, W and C represent the height, width and channel numbers of the image, respectively. First, the output features of four encoder blocks and bottle layer are denoted as $F_i \in \mathbb{R}^{\frac{H}{2^{i-1}} \times \frac{W}{2^{i-1}} \times C_i}$, where $i \in \{1, 2, 3, 4\}$, respectively. Then, for each level of MPCA block, $F_i$ and $F_{i+1}$ are used as inputs to accomplish cross-scale channel feature learning. The output of each MPCA block will go through a corresponding FSA block for spatial and frequency dual domain feature learning. Finally, a prediction head is used to get the final segmentation result $O$.
\subsection{Multi-scale Progressive Channel Attention block}

The detailed structure of MPCA block (using MPCA 1 as an example) is illustrated in Fig \ref{fig2} (b). Each MPCA block has two inputs, i.e., the current and next encoder output feature maps $F_i$ and $F_{i+1}$, respectively. Then the $F_i$ and $F_{i+1}$ will go through two independent Global Average Pooling (GAP) and $1 \times 1$ convolutional layers for dimension reduction and feature extraction before being concatenated. Finally, another $1 \times 1$ convolutional layer will be used to fuse the two feature maps, obtaining the multi-scale channel attention map $A$. It can be defined by the following Eq. \ref{eq1} and \ref{eq2}:
\begin{equation}
\label{eq1}
    F_{i(+1)}^{'}=\text{Conv}_{1\times1}(\text{GAP}(F_{i(+1)}))
\end{equation}
\begin{equation}
\label{eq2}
    A=\sigma(\text{Conv}_{1\times1}(\text{Concate}(F_i^{'}, F_{i+1}^{'})))
\end{equation}
where $\text{Conv}_{1\times1}$ denotes the $1\times1$ convolutional layers and Concate denotes concatenation along channel dimension. $A$ fused the channel features from two adjacent scales and completed cross-scale information exchange. Next, we split $A$ into two parts, $A_i$ and $A_{i+1}$, and dimensions are consistent with $F_i$ and $F_{i+1}$. Subsequently, $F_i$ and $F_{i+1}$ will be multiplied by the corresponding $A_i$ and $A_{i+1}$ to obtain the weighted feature maps $\hat{F_i}$ and $\hat{F_{i+1}}$. We use a transposed convolutional layer to upsample the feature map $\hat{F_{i+1}}$ by a factor of 2 and match the channel number with $\hat{F_i}$, resulting in $\hat{F_{i+1}^{'}}$. Finally, we perform element-wise addition between $\hat{F_i}$ and $\hat{F_{i+1}^{'}}$ to achieve feature fusion and obtain $\hat{F_{i}^{'}}$, which serves as the output of the MPCA block and the input of the FSA block. The above process can be represented by the following Eq.\ref{eq3}, \ref{eq4} and \ref{eq5}. We will introduce the structure of FSA block in the next section.
\begin{equation}
\label{eq3}
    A_i, A_{i+1}=\text{Split}(A)
\end{equation}
\begin{equation}
\label{eq4}
    \hat{F_{i(+1)}}=F_{i(+1)} \otimes A_{i(+1)}
\end{equation}
\begin{equation}
\label{eq5}
    \hat{F_{i}^{'}}=\hat{F_i} \oplus \text{UP}(\hat{F_{i+1}})
\end{equation}
Here, $\otimes$ and $\oplus$ represent element-wise multiplication and addition, respectively. UP denotes the transposed convolutional layer.

\subsection{Frequency-Spatial Attention block}
The detailed structure of FSA block (using FSA 1 as an example) is illustrated in Fig \ref{fig2} (c). Given the output $\hat{F_{i}^{'}}\in \mathbb{R}^{H\times W \times C_1}$  of MPCA block, we initially convert it into the frequency domain using the 2D DFT as depicted in Eq. \ref{eq6}. Here, $f_{i}\in \mathbb{R}^{H\times W \times C_1}$ represents the frequency domain feature maps, $\hat{F_{i}^{'}(x,y)}$ denotes the pixel values of the original features in the spatial domain, H and W denote the height and width of the feature maps, respectively. 

\begin{equation}
\label{eq6}
    f_{i}(U, V) = \sum_{x=0}^{H-1} \sum_{y=0}^{W-1} \hat{F_{i}^{'}}(x, y) e^{-j2\pi\left(\frac{Ux}{H} + \frac{Vy}{W}\right)}
\end{equation}

Then, we separate the frequency features into high-frequency $f_{i}^{high}$ and low-frequency $f_{i}^{low}$. This is achieved with two masks, $M_{low}\in \mathbb{R}^{H\times W \times 1}$ and $M_{high}\in \mathbb{R}^{H\times W \times 1}$. For $M_{low}$, a square of side length $n$ is centered in the mask and assigned a value of $1$, while the remaining regions are assigned a value of 0. Then, we perform element-wise multiplication between $f_{i}$ and $M_{low}$ to obtain the low-frequency component $f_{i}^{low}$, as defined in Eq. \ref{eq7}. For $M_{high}$, the small square area is assigned a value of $0$ and the rest is assigned to $1$. We then perform element-wise multiplication between $f_{i}$ and $M_{high}$ to obtain the high-frequency component $f_{i}^{high}$, as depicted in Eq. \ref{eq8}. In Fig \ref{fig2} (c), red diagonal lines are used to visually represent the regions assigned a value of $0$.
\begin{equation}
\label{eq7}
    f_{i}^{low}=f_{i} \otimes M_{low}
\end{equation}
\begin{equation}
\label{eq8}
    f_{i}^{high}=f_{i} \otimes M_{high}
\end{equation}

Subsequently, for $f_{i}^{low}$, we apply an adaptive learnable global filter $N \in \mathbb{R}^{H\times W \times C_1}$ to perform filtering. Then, we reintegrate the original high-frequency feature $f_{i}^{high}$ with the filtered $f_{i}^{low'}$, resulting  in the spectrum $f_i^{'}$ after global learning and feature fusion, as depicted in Eq. \ref{eq9}. Following this, a 2D IDFT is employed, as defined in Eq. \ref{eq10}, to obtain the spatial domain feature $\hat{F_{i}^{''}}$. 
\begin{equation}
\label{eq9}
    f_{i}^{'}=f_{i}^{high} \oplus (f_{i}^{low} \otimes N)
\end{equation}
\begin{equation}
\label{eq10}
    \hat{F_{i}^{''}}(x, y) = \frac{1}{HW} \sum_{U=0}^{H-1} \sum_{V=0}^{W-1} f_{i}^{'}(U,V) e^{j2\pi\left(\frac{Ux}{H} + \frac{Vy}{W}\right)}
\end{equation}

Meanwhile, for the output $\hat{F_{i}^{'}}$ from MPCA, we also use the spatial attention (SA) \cite{woo2018cbam} within the spatial branch to facilitate feature learning. The final output of the FSA block can be obtained as follows: 
\begin{equation}
\label{eq11}
    \widehat{F_{i}}=\hat{F_{i}^{''}} \oplus SA(\hat{F_{i}^{'}})
\end{equation}

\subsection{Decoder blocks}
In each decoder block, we employ linear interpolation to upsample the feature maps by a factor of $2$. Subsequently, we concatenate the high-resolution feature map with the upsampled feature maps along the channel dimension. Then, the concatenated feature maps undergo two $3 \times 3$ convolutional layers to complete feature fusion and learning. Each convolutional layer followed by a ReLU activation function, enhancing the network's nonlinear learning capability. The entire process can be defined by Eq. \ref{eq12}. It is important to note that $D_1$ passes through a prediction head consisting of two $3 \times 3$ convolutional layers to obtain the final segmentation result $O$ $\in \mathbb{R}^{H\times W \times n}$, where H and W represent the height and width of the image, respectively, and n represents the number of classes.
\begin{equation}
\label{eq12}
    D_i=\text{Relu}(\text{Conv}(\text{Concate}(\widehat{F_{i}}, \text{UP}(\widehat{F}_{i+1})))) \quad i \in (1,2,3,4)
\end{equation}

\section{Experiments}

\subsection{Datasets}
\subsubsection{\textbf{ISIC-2018}\cite{codella2019skin}}
This dataset comprises 2594 training images, 100 validation images, and 1000 test images. Each image is captured using professional dermatoscopic equipment and annotated by experts. For our experiments, we use the official training, validation, and test sets provided by ISIC.

\subsubsection{\textbf{BUSI}\cite{al2020dataset}}
The Breast Ultrasound Image Dataset (BUSI) consists of 647 breast ultrasound images from different patients, including 437 images of benign tumors and 210 images of malignant tumors. For our experiments, we randomly divide them into training (487 images), validation (80 images), and test sets (80 images).

\subsubsection{\textbf{NKUT}\cite{zhou2024nkut}}NKUT is a 3D CBCT
dataset used for pediatric mandibular wisdom tooth germ segmentation and it consists of 133 CBCT scans. Each data is carefully annotated at the pixel level by two experts for the mandibular wisdom teeth (MWT), second molars (SM), and the surrounding alveolar bone (AB). In our experiment, we adopt the same preprocessing methods outlined in \cite{zhou2024nkut}. Finally, we randomly divided the acquired images into training set (17251 images), validation set (1450 images), and test set (2058 images).

\subsection{Implementation details}
In our experiments, we applied data augmentations, including random horizontal flipping, random vertical flipping and random rotation. To ensure fairness, all models were trained using the same strategies. We used the Adam \cite{loshchilov2018decoupled} optimizer with an initial learning rate of 0.0001, which was decayed using the “Poly” strategy. The number of epochs was set to 200 and the loss functions was a combination of cross-entropy and Dice loss \cite{milletari2016v}. The channel numbers for the four enconder blocks were set to $[64, 128, 256, 512]$. For all datasets, we resized the images and labels to $224 \times 224$. Model parameters were randomly initialized without leveraging any pre-trained weights. For all models, we  selected the weights achieving the top three IoU scores on the validation set for testing on the test set, and retain the best result. Our framework was implemented using Pytorch, all experiments were conducted on two NVIDIA GeForce RTX 3090 GPUs.

\subsection{Experimental results}
\subsubsection{\textbf{Results on ISIC-2018}}
Table \ref{tab1} shows the quantitative results on the ISIC-2018 dataset. It is evident that SF-UNet outperformed previous SOTA networks in terms of DSC and IOU on the test set, highlighting the robust multi-scale feature learning and boundary feature learning capabilities of SF-UNet.


\begin{figure*}[h]
\centerline{\includegraphics[width=\linewidth]{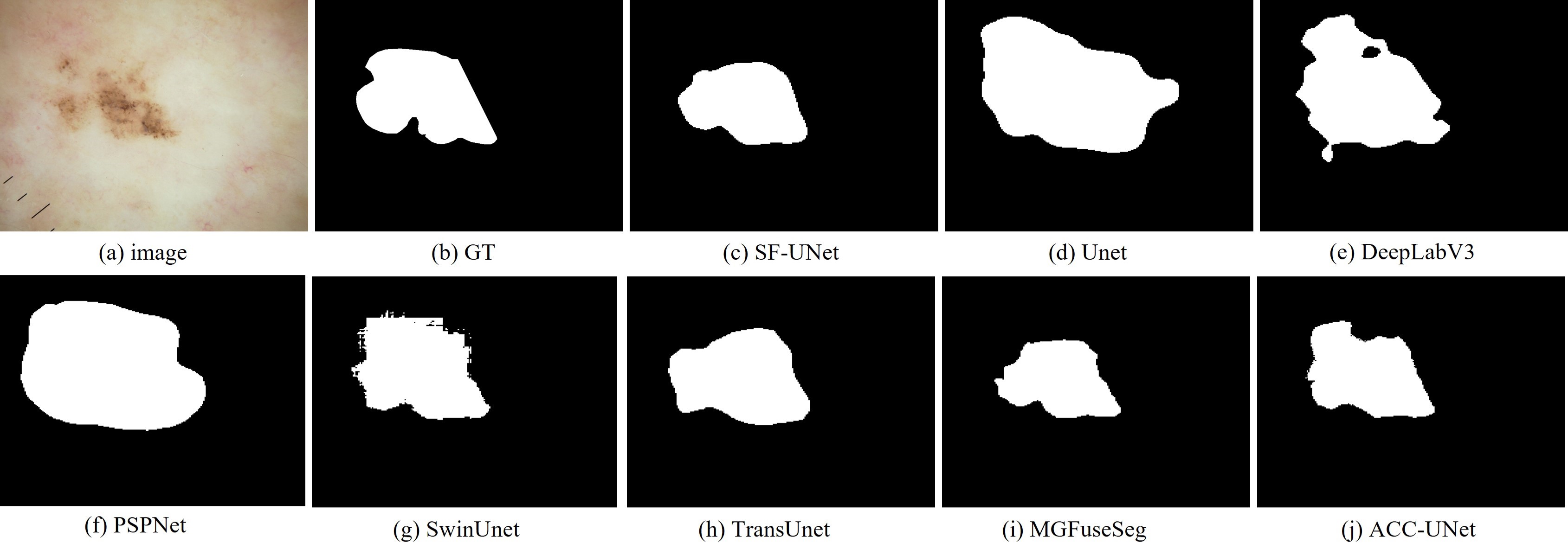}}
\vspace{-0.2cm}\caption{Qualitative results on ISIC-2018 dataset. (a) represents the original image and (b) represents the corresponding ground truth. Black represents the background, i.e., normal skin, while white represents lesions. It can be observed that SF-UNet achieves the best performance.}
\label{fig3}
\end{figure*}


\begin{figure*}[h]
\centerline{\includegraphics[width=\linewidth]{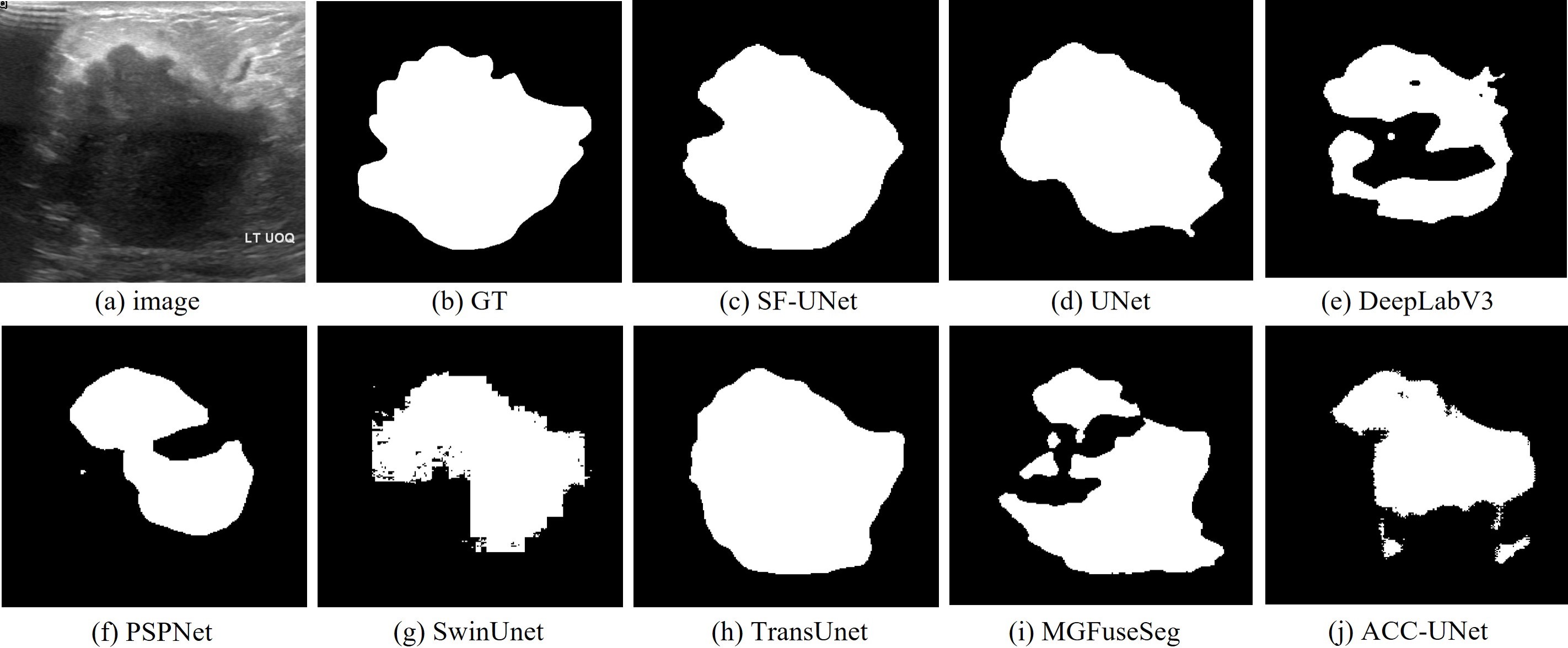}}
\vspace{-0.2cm}\caption{Qualitative results on BUSI dataset. (a) represents the original image and (b) represents the corresponding ground truth.  Black represents the background, while white represents tumors. It can be observed that SF-UNet achieves the best performance.}
\label{fig4}
\end{figure*}


\begin{figure*}[htbp]
\centerline{\includegraphics[width=\linewidth]{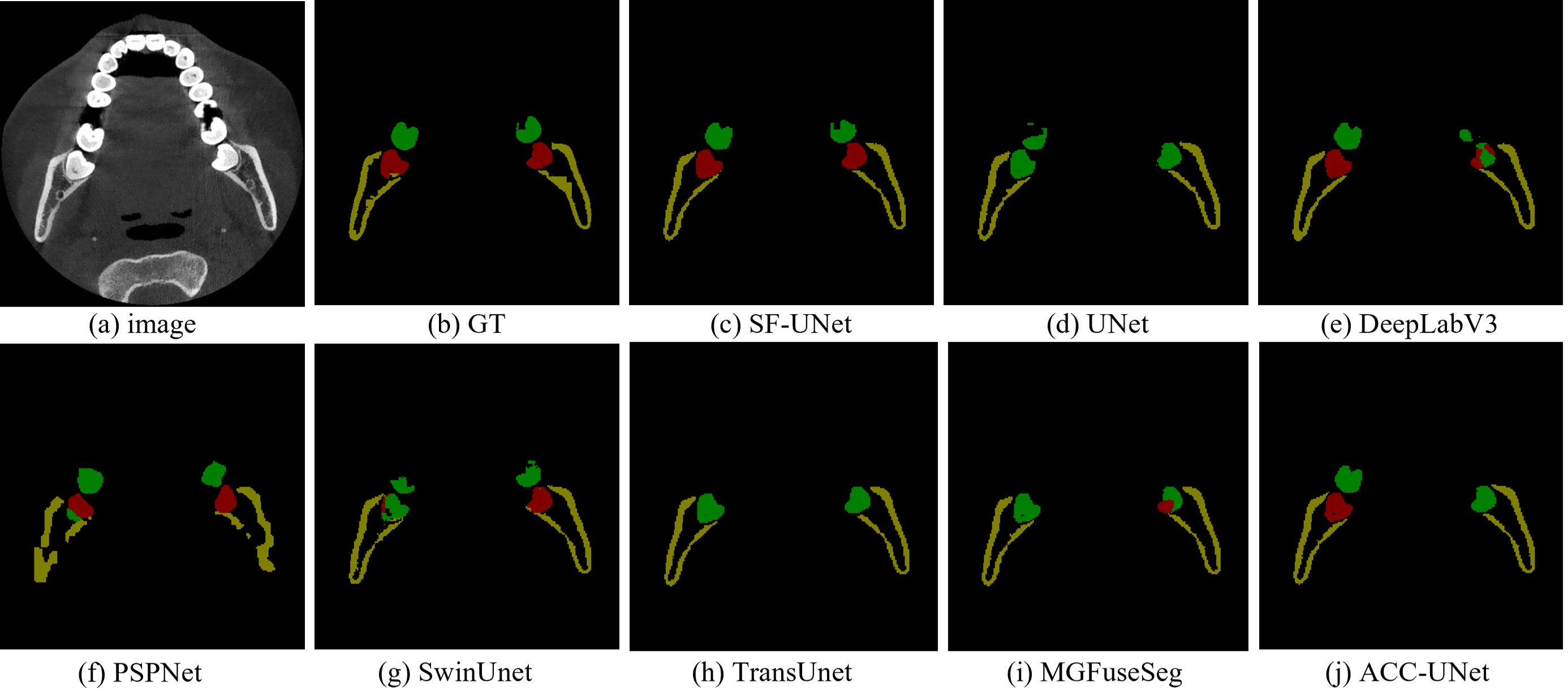}}
\vspace{-0.2cm}\caption{Qualitative results on NKUT dataset. (a) represents the original image and (b) represents the corresponding ground truth. Red represents Mandibular Wisdom Teeth (MWT), green represents Second Molars (SM), and yellow represents Alveolar Bone (AB).}
\label{fig5}
\end{figure*}

\begin{table}[htbp]
\centering
\small
\caption{the quantitative experimental results on the ISIC-2018 dataset, params denotes the numbers of Parameter. The bold results indicate the best performance, and underlined results indicate the second best.}
\label{tab1}
\begin{adjustbox}{width=\columnwidth} 
\setlength{\extrarowheight}{2.2pt} 
\begin{tabular}{|c|c|c|c|c|c|}
\hline
Dataset & Model & $\#$Params & Source \& Year & DSC(\%)$\uparrow$ & IOU(\%)$\uparrow$ \\ \hline
\multirow{8}{*}{ISIC-2018} & UNet\cite{ronneberger2015u} & 24.89M &MICCAI 2015  & 87.93 & 80.34 \\ 
 &DeepLabV3+\cite{vaswani2017attention}  & 20.15M &ECCV 2018  & 88.17 &  80.67\\  
 &PSPNet\cite{zhao2017pyramid}  & 16.16M &CVPR 2017  & 87.81 &80.26  \\ 
 &TransUNet\cite{chen2021transunet}  & 92.23M &ArXiv 2021  & \underline{88.27} & \underline{81.27} \\
 &Swin-UNet\cite{cao2022swin}  & 27.15M &ECCV 2022  & 87.74 &  80.13\\
 &ACC-UNet\cite{ibtehaz2023acc}  & 16.77M &MICCAI 2023  & 87.81 &  80.45\\
 &MGFuseSeg\cite{xu2023mgfuseseg}  & 22.27M &BIBM 2023  & 87.76 &  80.43\\ \cline{2-6}
 &SF-UNet (ours)  & 28.89M &/  & \textbf{88.46}&  \textbf{81.34}\\ \hline
 
\end{tabular}
\end{adjustbox}
\end{table}

Fig \ref{fig3} presents the qualitative results on the ISIC-2018 dataset. It is visually apparent that SF-UNet produces segmentation results with the smoothest and clearest boundaries, closely resembling the ground truth. In contrast, The segmentation results of other networks display problems such as loss of texture details or boundary confusion. The results highlights the effectiveness of SF-UNet and preserves high-frequency features from a frequency-domain perspective to generate accurate segmentation results.

\subsubsection{\textbf{Results on BUSI}}

Table \ref{tab2} displays the quantitative results on the BUSI dataset. It can be observed that SF-UNet achieved the best performance in terms of HD95, DSC, and IOU, surpassing the previous SOTA networks. Compared to the second-best ACC-UNet, it mproves the results by 0.64\%, 0.42\%, and 0.76\% respectively. This demonstrates that SF-UNet exhibits high robustness, achieving excellent generalization performance even on small datasets.

\begin{table}[h]
\centering
\small
\caption{the quantitative results on the BUSI dataset, params denotes the numbers of Parameter. The bold results indicate the best performance, and underlined results indicate the second best.}
\label{tab2}
\begin{adjustbox}{width=\columnwidth} 
\setlength{\extrarowheight}{2.2pt} 
\begin{tabular}{|c|c|c|c|c|c|}
\hline
Dataset & Model & $\#$Params & HD95$\downarrow$ & DSC(\%)$\uparrow$ & IOU(\%)$\uparrow$ \\ \hline
\multirow{8}{*}{BUSI} & UNet\cite{ronneberger2015u} & 24.89M &101.79  & 73.27 & 64.30 \\ 
 &DeepLabV3+\cite{vaswani2017attention}  & 20.15M &101.13  & 74.74 &  65.17\\  
 &PSPNet\cite{zhao2017pyramid}  & 16.16M &80.67  & 72.82 &63.24  \\ 
 &TransUNet\cite{chen2021transunet}  & 92.23M &97.91  & 74.89 & 66.35 \\
 &Swin-UNet\cite{cao2022swin}  & 27.15M &112.34  & 66.66 &  56.42\\
 &ACC-UNet\cite{ibtehaz2023acc}  & 16.77M &\underline{80.17}  & \underline{75.64} &  \underline{66.44}\\
 &MGFuseSeg\cite{xu2023mgfuseseg}  & 22.27M &113.62 & 67.69 &  57.52\\ \cline{2-6}
 &SF-UNet (ours)  & 28.89M &\textbf{79.53}  & \textbf{76.06}&  \textbf{67.20}\\ \hline
 
\end{tabular}
\end{adjustbox}
\end{table}

Fig \ref{fig4} presents the qualitative results and it is evident that among all networks, SF-UNet achieves segmentation results with more accurate and clear boundaries. Other networks display issues such as boundary missing or excessive smoothness, along with inaccurate lesion detection. This further demonstrates the importance of the MPCA and FSA blocks, which enable the network to accurately learn texture features of lesions at multiple scales and precisely capture high-frequency boundary features, leading to better segmentation performance.

\subsubsection{\textbf{Results on NKUT}}
 Table \ref{tab3} presents the quantitative results on the NKUT dataset. From the results, SF-UNet achieved the best performance in terms of AB and average performance, and achieved the second best performance in terms of MWT and SM. For TransUNet, despite outperforming SF-UNet in the segmentation of MWT and SM, it has two limitations. Firstly, its parameters (92.23M) are significantly higher than SF-UNet. Secondly, its performance in segmenting AB is subpar, indicating inadequate multi-scale and boundary information learning capabilities. Its high IOU in MWT and SM may stem from aggressive classification of ambiguous boundaries between teeth and bones as teeth, compromising AB accuracy to enhance MWT and SM accuracy. Therefore, considering these aspects comprehensively, SF-UNet remains competitive.
\begin{table}[H]
\centering
\small
\caption{the quantitative experimental results on the NUKT dataset, We chose IOU as the evaluation metric. MWT represents the mandibular wisdom tooth germ, SM indicates the second mandibular molar, and AB represents the alveolar bone. The bold results indicate the best performance, and underlined results indicate the second best.}
\label{tab3}
\begin{adjustbox}{width=\columnwidth} 
\setlength{\extrarowheight}{2.2pt} 
\begin{tabular}{|c|c|c|c|c|c|}
\hline
Dataset & Model & MWT & SM & AB & Mean\\ \hline
\multirow{8}{*}{NKUT} & UNet\cite{ronneberger2015u} & 84.61 &81.61  & 53.56 & 73.26 \\ 
 &DeepLabV3+\cite{vaswani2017attention}  & 85.54 &84.86  & 52.97 &  74.46\\  
 &PSPNet\cite{zhao2017pyramid}  & 83.91 &82.17  & 42.37 &69.48  \\ 
 &TransUNet\cite{chen2021transunet}  & \textbf{88.37} &\textbf{83.76}  & 51.79 & \underline{74.64} \\
 &Swin-UNet\cite{cao2022swin}  & 81.12 &76.78  & \underline{55.01} &  70.97\\
 &ACC-UNet\cite{ibtehaz2023acc}  & 85.61 &83.07  & 53.43 &  74.04\\
 &MGFuseSeg\cite{xu2023mgfuseseg}  & 86.02 &81.09  & 53.41 &  73.51\\ \cline{2-6}
 &SF-UNet (ours)  & \underline{86.15} &\underline{82.59}  & \textbf{56.27}&  \textbf{75.00}\\ \hline
 
\end{tabular}
\end{adjustbox}
\end{table}

Fig \ref{fig5} illustrates the qualitative results on the NKUT dataset. We can observe that SF-UNet can effectively delineate MWT, SM, and AB, demonstrating clear boundaries between teeth and bone. Despite the extreme morphological similarity between the MWT and SM in the image,  SF-UNet accurately distinguishes between them. This demonstrates that SF-UNet effectively learns deep and high-level features. In contrast, other networks display various issues, including missing teeth, ambiguous teeth, omitted bone structures, or unclear boundaries between teeth and bone.

\subsection{Ablation studies}
We conducted ablation studies of SF-UNet on the BUSI dataset to validate the effectiveness of the proposed MPCA and FSA blocks. In the second row of Tabel \ref{tab4}, it can be observed that integrating the FSA block into the backbone UNet leads to improvements in both DSC and IOU, with increases of 0.83\% and 0.81\% , respectively. Meanwhile, the parameters are only increased by 0.05M, demonstrating the the lightweight and effectiveness of the FSA block. In the third row of the Tabel \ref{tab4}, the ntegration of only the MPCA block to the backbone results in significant enhancements in both DSC and IOU, increasing by 2.24\% and 3.17\%, respectively. This underscores the multi-scale learning capability of MPCA, substantially bolstering the network's robustness and effectiveness in addressing multi-scale lesion segmentation challenges. In the last row of Table \ref{tab4}, upon integrating both MPCA and FSA into the backbone, the SF-UNet is formed, achieving the best performance in both DSC and IOU, with improvements of 2.79\% and 2.9\%, respectively, compared to the backbone UNet. This demonstrates the complementary nature of MPCA and FSA, indicating that both of them are indispensable.
\begin{table}[h]
\centering
\small
\caption{This table shows the ablation experiment results of SF-UNet on the BUSI dataset, where $\times$ indicates the block is not included, and $\checkmark$ indicates the block is included. We chose DSC and IOU as the evaluation metrics.}
\label{tab4}
\begin{adjustbox}{width=\columnwidth} 
\setlength{\extrarowheight}{1pt} 
\begin{tabular}{|c|c c|c|c|c|}
\hline
Backbone & MPCA & FSA & DSC & IOU & Params\\ \hline
\multirow{4}{*}{UNet\cite{ronneberger2015u}} & $\times$ & $\times$ &73.27  & 64.30 & 24.89M \\  
 &$\times$ & $\checkmark$ &74.10  & 65.11 &24.94M  \\ 
  &$\checkmark$  & $\times$ &75.51  & 67.47 &  28.84M\\ 
 &$\checkmark$  & $\checkmark$ &\textbf{76.06}  & \textbf{67.20}&  28.89M\\ \hline
 
\end{tabular}
\end{adjustbox}
\vspace{-0.6cm}
\end{table}

\section{Conclusion}
In this paper, we proposed SF-UNet for medical image segmentation, which consists of two blocks: MPCA and FSA blocks. We conducted extensive experiments on three public datasets, and the results showed that SF-UNet achieved the best performance, surpassing previous SOTA networks. Compared to other networks, SF-UNet demonstrated superior accuracy in segmenting textures and boundaries, as well as in distinguishing between various lesion types and sizes.

\section*{Acknowledgment}
This work is partially supported by the National Natural Science Foundation (62272248), the Natural Science Foundation of Tianjin (23JCZDJC01010, 23JCQNJC00010). 
\bibliographystyle{ieeetr}
\bibliography{ref}

\end{document}